\documentclass[twocolumn]{aastex631}

\newcommand{\qfit}{\texttt{q3dfit}}

\newcommand{\oi}{\hbox{[O$\,${\scriptsize I}]}}

\newcommand{\oiii}{\hbox{[O$\,${\scriptsize III}]}}

\newcommand{\nii}{\hbox{[N$\,${\scriptsize II}]}}
\newcommand{\sii}{\hbox{[S$\,${\scriptsize II}]}}
\newcommand{\ha}{\hbox{H$\alpha$}}
\newcommand{\hb}{\hbox{H$\beta$}}

\newcommand{\niiha}{log([N{\sc II}]/H$\alpha$)}
\newcommand{\oiiihb}{log([O{\sc III}]/\hb)}
\newcommand{\siiha}{log([S{\sc II}]/\ha)}
\newcommand{\oiha}{log([O{\sc I}]/\ha)}
\newcommand{\oh}{log([O/H])+12}

\newcommand{\kms}{km\,s$^{-1}$} 
 
\newcommand{\eden}{cm$^{-3}$}

\newcommand{\ergs}{erg s$^{-1}$ }
\newcommand{\myr}{M$_\odot$~yr$^{-1}$} 

\newcommand{\logohb}{log(\oiii/\hb) }
\renewcommand{\deg}{\ensuremath{^{\circ}}}
\newcommand{\ferg}{erg s$^{-1}$ cm$^{-2}$ }
\newcommand{\surff}{erg s$^{-1}$ cm$^{-2}$ \AA$^{-1}$arcsec$^{-2}$}

\shorttitle{Clumpy star formation and shocks in red quasar}
\shortauthors{Vayner et al.}
\graphicspath{{./}{figures/}}

\begin{document}

\title{First results from the JWST Early Release Science Program Q3D: Ionization cone, clumpy star formation and shocks in a $z=3$ extremely red quasar host}

\correspondingauthor{Andrey Vayner}
\email{avayner1@jhu.edu}

\author[0000-0002-0710-3729]{Andrey Vayner}
\affiliation{Department of Physics and Astronomy, Bloomberg Center, Johns Hopkins University, 3400 N. Charles St., Baltimore, MD 21218, USA}

\author[0000-0001-6100-6869]{Nadia L. Zakamska}
\affiliation{Department of Physics and Astronomy, Bloomberg Center, Johns Hopkins University, 3400 N. Charles St., Baltimore, MD 21218, USA}

\author[0000-0001-7572-5231]{Yuzo Ishikawa}
\affiliation{Department of Physics and Astronomy, Bloomberg Center, Johns Hopkins University, 3400 N. Charles St., Baltimore, MD 21218, USA}

\author[0000-0002-4419-8325]{Swetha Sankar}
\affiliation{Department of Physics and Astronomy, Bloomberg Center, Johns Hopkins University, 3400 N. Charles St., Baltimore, MD 21218, USA}

\author[0000-0003-2212-6045]{Dominika Wylezalek}
\affiliation{Zentrum für Astronomie der Universität Heidelberg, Astronomisches Rechen-Institut, Mönchhofstr 12-14, D-69120 Heidelberg, Germany}

\author[0000-0002-1608-7564]{David S. N. Rupke}
\affiliation{Department of Physics, Rhodes College, Memphis, TN 38112, USA}

\author[0000-0002-3158-6820]{Sylvain Veilleux}
\affiliation{Department of Astronomy and Joint Space-Science Institute, University of Maryland, College Park, MD 20742, USA}

\author[0000-0002-6948-1485]{Caroline Bertemes}
\affiliation{Zentrum für Astronomie der Universität Heidelberg, Astronomisches Rechen-Institut, Mönchhofstr 12-14, D-69120 Heidelberg, Germany}

\author[0000-0003-2405-7258]{Jorge K. Barrera-Ballesteros}
\affiliation{Instituto de Astronomía, Universidad Nacional Autónoma de México, AP 70-264, CDMX 04510, Mexico}

\author[0000-0001-8813-4182]{Hsiao-Wen Chen}
\affiliation{Department of Astronomy \& Astrophysics, The University of Chicago, 5640 South Ellis Avenue, Chicago, IL 60637, USA}

\author{Nadiia Diachenko}
\affiliation{Department of Physics and Astronomy, Bloomberg Center, Johns Hopkins University, 3400 N. Charles St., Baltimore, MD 21218, USA}

\author[0000-0003-4700-663X]{Andy D. Goulding}
\affiliation{Department of Astrophysical Sciences, Princeton University, 4 Ivy Lane, Princeton, NJ 08544, USA}

\author[0000-0002-5612-3427]{Jenny E. Greene}
\affiliation{Department of Astrophysical Sciences, Princeton University, 4 Ivy Lane, Princeton, NJ 08544, USA}

\author[0000-0003-4565-8239]{Kevin N. Hainline}
\affiliation{Steward Observatory, University of Arizona, 933 North Cherry Avenue, Tucson, AZ 85721, USA}

\author{Fred Hamann}
\affiliation{Department of Physics \& Astronomy, University of California, Riverside, CA 92521, USA}

\author[0000-0001-8813-4182]{Timothy Heckman}
\affiliation{Department of Physics and Astronomy, Bloomberg Center, Johns Hopkins University, Baltimore, MD 21218, USA}

\author[0000-0001-9487-8583]{Sean D. Johnson}
\affiliation{Department of Astronomy, University of Michigan, Ann Arbor, MI 48109, USA}

\author{Hui Xian Grace Lim}
\affiliation{Department of Physics, Rhodes College, Memphis, TN 38112, USA}

\author[0000-0003-3762-7344]{Weizhe Liu}
\affiliation{Department of Astronomy, Steward Observatory, University of Arizona, Tucson, AZ 85719, USA}

\author[0000-0003-0291-9582]{Dieter Lutz}
\affiliation{Max-Planck-Institut für Extraterrestrische Physik, Giessenbachstrasse 1, D-85748 Garching, Germany}

\author[0000-0001-6126-5238]{Nora Lützgendorf}
\affiliation{European Space Agency, Space Telescope Science Institute, Baltimore, Maryland, USA}

\author[0000-0002-1047-9583]{Vincenzo Mainieri}
\affiliation{European Southern Observatory, Karl-Schwarzschild-Straße 2, D-85748 Garching bei München, Germany}

\author{Ryan McCrory}
\affiliation{Department of Physics, Rhodes College, Memphis, TN 38112, USA}

\author[0009-0007-7266-8914]{Grey Murphree}
\affiliation{Department of Physics, Rhodes College, Memphis, TN 38112, USA}
\affiliation{Institute for Astronomy, University of Hawai'i, 2680 Woodlawn Dr., Honolulu, HI, 96822, USA}

\author[0000-0001-5783-6544]{Nicole P. H. Nesvadba}
\affiliation{Université de la Côte d'Azur, Observatoire de la Côte d'Azur, CNRS, Laboratoire Lagrange, Bd de l'Observatoire, CS 34229, Nice cedex 4 F-06304, France}

\author[0000-0002-3471-981X]{Patrick Ogle}
\affiliation{Space Telescope Science Institute, 3700, San Martin Drive, Baltimore, MD 21218, USA}

\author[0000-0002-0018-3666]{Eckhard Sturm}
\affiliation{Max-Planck-Institut für Extraterrestrische Physik, Giessenbachstrasse 1, D-85748 Garching, Germany}

\author{Lillian Whitesell}
\affiliation{Department of Physics, Rhodes College, Memphis, TN 38112, USA}

\begin{abstract}
Massive galaxies formed most actively at redshifts $z=1-3$ during the period known as `cosmic noon.' Here we present an emission-line study of an extremely red quasar SDSSJ165202.64+172852.3 host galaxy at $z=2.94$, based on observations with the Near Infrared Spectrograph (NIRSpec) integral field unit (IFU) on board JWST. We use standard emission-line diagnostic ratios to map the sources of gas ionization across the host and a swarm of companion galaxies. The quasar dominates the photoionization, but we also discover shock-excited regions orthogonal to the ionization cone and the quasar-driven outflow. These shocks could be merger-induced or -- more likely, given the presence of a powerful galactic-scale quasar outflow -- these are signatures of wide-angle outflows that can reach parts of the galaxy that are not directly illuminated by the quasar. Finally, the kinematically narrow emission associated with the host galaxy presents as a collection of 1 kpc-scale clumps forming stars at a rate of at least 200 $M_{\odot}$ yr$^{-1}$. The ISM within these clumps shows high electron densities, reaching up to 3,000 cm$^{-3}$ with metallicities ranging from half to a third solar with a positive metallicity gradient and V band extinctions up to 3 magnitudes. The star formation conditions are far more extreme in these regions than in local star-forming galaxies but consistent with that of massive galaxies at cosmic noon. JWST observations reveal an archetypical rapidly forming massive galaxy undergoing a merger, a clumpy starburst, an episode of obscured near-Eddington quasar activity, and an extremely powerful quasar outflow simultaneously.
\end{abstract}


\section{Introduction} 
\label{sec:intro}

The most massive galaxies in the present-day universe are passively evolving elliptical galaxies devoid of a significant amount of molecular gas and showing little star formation in the last 10 Gyr, with all present-day star formation concentrated in lower-mass disks. In contrast, at cosmic noon when the universe was 2.2-4.3 Gyr old, or at $z=1.5-3$ -- the epoch known as `cosmic noon' -- the highest star formation rate galaxies were also the most massive \citep{cowi96, font09, conr09, hall18}. This epoch also corresponds to the peak star formation rate density and the black hole accretion density \citep{boyl98, mada14}, with $\sim 50\%$ of all stars observed today formed at $z=1.5-3$. Massive gas-rich galaxies at high redshifts undergo the most rapid galactic evolution and display the highest measured star formation rates, triggered by secular processes \citep{fors20} and mergers \citep{enge10}. Morphologically, this starburst activity proceeds in turbulent, thickened disks \citep{law12} or in multiple clumps \citep{fors09, mene13, iono16}. Exactly how these extremely active and morphologically disturbed galaxies become quenched and settle on the passively evolving sequence of smooth elliptical galaxies remains a major open question in galaxy formation, with secular processes and feedback from star formation and quasars potentially playing a role \citep{nogu98, hopk06, mart09, ceve10}. 

Dust-reddened quasars may be a particularly interesting population as they are expected to be transitional sources, where quasar-driven winds may be in the act of clearing out the host galaxy and quenching star formation \citep{sand88, hopk06}. Objects with such properties have been long sought, and a particularly promising population is that of extremely red quasars (ERQs; \citealt{ross15, hama17}). These objects are selected from WISE \citep{wrig10} and SDSS \citep{eise11} based on their extremely red color, $i-W3 > 4.6$ AB mag, and highly unusual rest-frame ultraviolet emission-line profiles. Follow-up studies of this population have established that these quasars are uniquely associated with the highest velocity widths and blueshifts of the \oiii\ $\lambda$4959, 5007\AA\ emission lines, indicative of powerful outflows \citep{zaka16b, perr19} which have now been demonstrated to extend on galaxy-wide scales \citep{vayner21a}. With bolometric luminosities exceeding 10$^{47}$ \ergs, these quasars are likely accreting at or beyond the Eddington limit. They may be associated with the long-sought `blow-out' phase of the quasar evolution, marking them ideal test subjects to understand a critical phase of early galaxy evolution associated with black hole growth and feedback. 

SDSS~J165202.64+172852.3 (J1652 hereafter) is one such ERQ at $z = 2.94$ with $i - W3=5.4$ AB mag. X-ray observations directly confirm intervening obscuration with near Compton-thick column densities \citep{goul18a, Ishikawa21}. A bolometric luminosity of $5 \times 10^{47}$ \ergs\ is estimated from the directly observed infrared WISE flux \citep{wrig10, goul18b, perr19}. \citet{alex17} and \citet{hwan18} classify the system as a radio-intermediate source, but there is no evidence for extended radio emission suggesting the presence of powerful jets. Hubble Space Telescope ({\it HST}) observations with Wide Field Camera 3 (WFC3), corresponding to the rest-frame $B$-band, suggest that the quasar is hosted by a massive galaxy with $M_*/M_{\odot}\sim11.4 - 12.4$ which is undergoing a major merger and exhibits a tidal feature in the western direction \citep{zaka19}. Despite the long-standing paradigm in which the red quasar phase is strongly associated with major mergers \citep{hopk06,glik15}, J1652 is unusual in demonstrating these signatures within the ERQ population \citep{zaka19}.

{\it JWST} \citep{Gardner06} observed J1652 with the Near-Infrared Spectrograph (NIRSpec; \citealt{Jakobsen22}) using the integral field unit (IFU) mode on 2022-07-15 as part of the Early Release Science Program ``Imaging Spectroscopy of Quasar Hosts with {\it JWST}'' \citep{Wylezalek22}; this was one of the first NIRSpec-IFU observations obtained by the observatory. NIRSpec reveals extended ionized gas emission across the entire field as traced by the ionized \oiii\ with at least three interacting companion galaxies \citep{Wylezalek22}. The companion galaxies are estimated to have $M_*>10^{9.8}M_{\odot}$ with individual velocities ranging between $[-450,+820]$ \kms. Extended \oiii\ associated with the {\it HST} tidal tail has also been detected at velocities of up to $v\sim800$ \kms. The density of companions and their extreme kinematics suggest that the dark-matter halo of the massive host of J1652 may itself be undergoing a major merger, and therefore the quasar may be tracing one of the densest and most actively forming cosmic knots at $z\sim3$.

J1652 was already known to exhibit $\sim$ 3000 km s$^{-1}$ \oiii\ velocities based on both the spatially integrated \citep{alex18} and spatially resolved \citep{vayner21a} ground-based near-IR IFU observations. Gemini Near-Infrared Integral Field Spectrometer (NIFS) adaptive-optics observations show that the region of extremely high-velocity dispersion of \oiii\ extends out to at least 4 kpc from the nucleus in the Southern direction \citep{vayner21a}. In a companion paper (Vayner et al. 2023b - in prep.), we present the in-depth study of the quasar-driven outflow, its geometry, and its energetics from the {\it JWST} NIRSpec data. The high spatial and spectral resolution of the {\it JWST} NIRSpec observations allows us to disentangle the emission associated with the outflow from that of the host galaxy and the companion galaxies based on the gas kinematics. We confirm the presence of the highly blue-shifted outflow with outflow velocities up to 1000 \kms\ on kpc scales (Vayner et al. 2023b - in prep.) and additionally reveal the red-shifted component of the outflow towards the North-East ($v = 450$ \kms\ and $\sigma_v$ = 500 \kms; \citealt{Wylezalek22}). The distribution of the velocity width of \oiii\ components in J1652 is bimodal, with a minimum at $\sigma_v=300$ \kms. The gas with smaller velocity dispersion is likely in dynamical equilibrium with the host galaxy of J1652 or with companion galaxies, whereas the gas with higher velocity dispersions is likely not confined by the gravitational potential of the galaxy and is in an outflow. 

Here we present an in-depth look at the emission-line ratio maps in J1652. These measurements will inform us about the sources of ionization, the geometry of ionization and outflows, the morphology of the host and the companions, the distribution of the star formation in the host galaxy, and about the physical origin of any shocked emission. We summarize observations, data reduction, and point-spread function subtraction in Section \ref{sec:obs}. We present the measurements of emission lines and line ratios in Section \ref{sec:analysis}. We discuss the origin of the features we see in the ionization maps in Section \ref{sec:discussion}, and we conclude in Section \ref{sec:conclusions}. We use a three-dimensional flat geometry of $\Lambda$CDM cosmological model with H$_{o}$ = 69.6 km s$^{-1}$ Mpc$^{-1}$, $\Omega_{M}$ = 0.286, and $\Omega_{\Lambda}$ = 0.714 to measure all distances \citep{bennett2014}. While we use vacuum emission line wavelengths for fitting as appropriate for space-based data, we use air wavelengths in line identification following a long-standing convention.

\section{Data} 
\label{sec:obs}

\subsection{Observational design and data reductions}

NIRSpec observations were set up with the G235H grating in combination with the F170LP filter, resulting in wavelength coverage 1.65$-$3.15 \micron\ and spectral resolution 85-150 \kms. There were no dedicated acquisition or verification exposures; the source was acquired and observed using the absolute pointing accuracy of the observatory. We used the ``NRSIRS2" readout mode for an effective exposure time per integration of 1823.6s, and a total on-source exposure time of 16412.5 s. We used the 9-point dither pattern (which allows offsets by half-integer number of pixels) to improve the spatial sampling of the point-spread function (PSF). The field of view of NIRSpec is 3\arcsec $\times$3\arcsec, corresponding to the physical scale of $\sim 25\times 25$ kpc$^2$ at the redshift of our source ($z=2.94$). However, due to our selected dithering pattern, we are able to detect extended emission in a slightly larger field of view; however, the edge spaxels have higher noise values due to a smaller number of exposures. Our final field of view is approximately 4\arcsec $\times$4\arcsec. At the first science dither position, we took a single exposure with all the micro-shutter assembly (MSA) closed to inspect and remove any light leakage from bright objects in the NIRSpec instrument field of view and to inspect and remove any background light from failed open shutters. 

We reduced the data using Space Telescope's {\it JWST} pipeline version 1.8.2 \footnote{\url{https://github.com/spacetelescope/jwst}}. The first stage performs standard infrared detector reduction on uncalibrated files (i.e, dark current subtraction, data quality flagging, bias subtraction, and a first iteration of cosmic ray removal) to produce rate files. These files are then fed to the second stage of the pipeline, which assigns a world coordinate system per frame, background subtracts, flat-fields, flux calibrates, and converts the 2D spectra into a 3D data cube via the ``cube build" routine. We use the ``emsm" routine instead of the standard ``drizzle" method when extracting the data from 2D to 3D \footnote{\url{https://jwst-pipeline.readthedocs.io/en/latest/jwst/cube_build/main.html##algorithm}}. The ``emsm" improved the oscillating spectral pattern in the point source spectrum compared to the ``drizzle" method at the cost of minor degradation in the spatial resolution. An additional procedure is implemented in this step to flag and subtract imprints produced by the open NIRSpec micro-shutters and clip bad pixels; however, we had to skip the imprint subtraction step as that increased the noise level in the final produced data cubes, likely because the ``leakcal" exposure was only taken at a single dither position. At the time of data reduction, the flat-field calibration files contained placeholder values; hence no proper flat fielding was done on the data. We skipped the third stage of the pipeline that combines the different exposures taken at different dither positions due to issues with the outlier detection step that kept a significant amount of bad pixels in the final data cube and masked a large portion of the quasar emission. Instead, we opted to use an in-house script based on the Python ``Reproject" \footnote{\url{https://pypi.org/project/reproject/}} package \citep{thomas_robitaille_2023_7584411}, and the flux-conserving ``reproject interp" routine together with the \textit{astropy} sigma clip routine to align the different dither positions produced by ``spec2pipeline" and remove significant flux outliers. Our custom pipeline projects the data cubes onto a 0.05\arcsec spatial grid and a spectral sampling of 0.0396 \micron. The final data cube has a wavelength range of 1.65 \micron -- 3.176 \micron. We apply the same procedure to the variance cubes produced by the second stage of the pipeline and similarly combine them to produce a final combined variance cube for the data. We applied the same reduction analysis to the commissioning standard star observations with NIRSpec of TYC 4433-1800-1 (PID 1128) to flux calibrate our data. We achieve a final 2$\sigma$ flux sensitivity of 2.87$\times10^{-19}$ \surff\ and AB magnitude/arcsec$^{2}$ of 22.45 at 1.995 \micron, near redshifted \oiii\ 5007 \AA\ and 2.45$\times10^{-19}$ \surff\ and AB magnitude/arcsec$^{2}$ of 22.09 at 2.549 \micron, near redshifted \ha. Our 2$\sigma$ surface brightness sensitivities are at-least a factor of two higher than the measured JWST background at 2-3 \micron\ \citep{Rigby23}, hence our observations are limited by the detector noise and systematics of the NIRSpec IFU.

There are still some challenges associated with the NIRSpec integral field unit data reduction. The full width at half maximum (FWHM) of the simulated point-spread function (PSF) for the Near-infrared Camera (NIRCam) of {\it JWST} is about 0.065\arcsec\ at 2 \micron\footnote{\url{https://jwst-docs.stsci.edu/jwst-near-infrared-camera/nircam-performance/nircam-point-spread-functions}}, and this size is primarily determined by the diffraction limit of the telescope. The 0.1\arcsec\ native size of the NIRSpec spaxel is about 3 times larger than would be required to Nyquist-sample this PSF. Our observations employ a 9-point half-integer dither pattern which is used to improve the PSF sampling, with the final cube projected onto an 0.05\arcsec\ grid. But the data are still undersampled, both spatially and spectrally, which results in ``wiggle" artifacts in the final spectrum. The ``wiggles" are somewhat reduced as we take spectra in wider apertures and use the ``emsm" method to extract the spectra from 2D to 3D, but of course that comes at the cost of degrading the effective spatial resolution. We achieve a final spatial resolution of approximately 200 mas.

\subsection{PSF subtraction}

A major challenge in studying the extended emission around quasars is the bright central quasar that outshines the faint host galaxy, sometimes by several orders of magnitude. Subtracting the quasar from imaging data to reveal the host galaxy requires exquisite knowledge of the PSF, either from simulations (e.g., as was done by \citealt{zaka19}) or from observations of stars (e.g., \citealt{glik15, mech16}). Subtracting the PSF from an integral-field unit data cube using theoretical models or samples of standard stars is more challenging because such data sets are rarely available. Some PSF subtraction procedures \citep{vayner16, rupk17} take advantage of the difference in the spectra between the quasar and the host. They construct a data-driven PSF using a wavelength region where the quasar dominates (e.g., the broad-line region of a permitted emission line) and then generate a cube by scaling the flux of the PSF in accordance with the quasar spectrum. 

In the NIRSpec data, the situation is further compounded by the PSF whose shape and size vary as a function of wavelength since it is primarily diffraction-limited. In ground-based data, for low-Strehl adaptive optics (AO) or seeing-limited PSF, its size variation over the wavelength range is, at most, a minor concern, whereas, for our data, the size of the PSF varies with wavelength. 

We use \qfit\footnote{\url{https://q3dfit.readthedocs.io/en/latest/index.html}} to model and remove the quasar PSF to reveal the faint extended emission. \qfit\ is a \texttt{Python}-based software, adapted from \texttt{IFSFIT} \citep{ifsfit2014, rupk15}, for scientific analysis of {\it JWST} integral field spectroscopy of quasars and their host galaxies (Rupke et al. in prep). \qfit\ works in three steps. First, it removes the central quasar PSF to produce a PSF-subtracted datacube containing the faint host galaxy emission. Next, the decomposed host galaxy emission is carefully fit with a combination of continuum, emission lines, and absorption lines. Finally, \qfit\ takes the best-fit emission line, and continuum outputs to produce maps and other science products. 

\qfit\ performs maximal-contrast subtraction of the quasar PSF by taking advantage of the spectral differences between quasars and their host galaxies. \qfit\ extracts the quasar spectrum using the brightest spaxel; specifically for J1652, we use a 2-pixel radius (0.1\arcsec) centered on the brightest spaxel to extract a quasar-dominated spectrum. This quasar spectrum is then fit by scaling the spectrum using a combination of multiplicative and additive polynomials and exponential functions across the NIRSpec FOV, which is then subtracted to reveal the faint extended emission corresponding to the host galaxy. The multiplicative and additive polynomials are selected to fit and account for the variation in the shape of the NIRSpec PSF as a function of wavelength (Rupke et al. in-prep.).

\section{Emission line analysis and maps} 
\label{sec:analysis}

\subsection{Multi-Gaussian fitting}
\begin{figure*}
    \centering
    \includegraphics[width=0.9\textwidth]{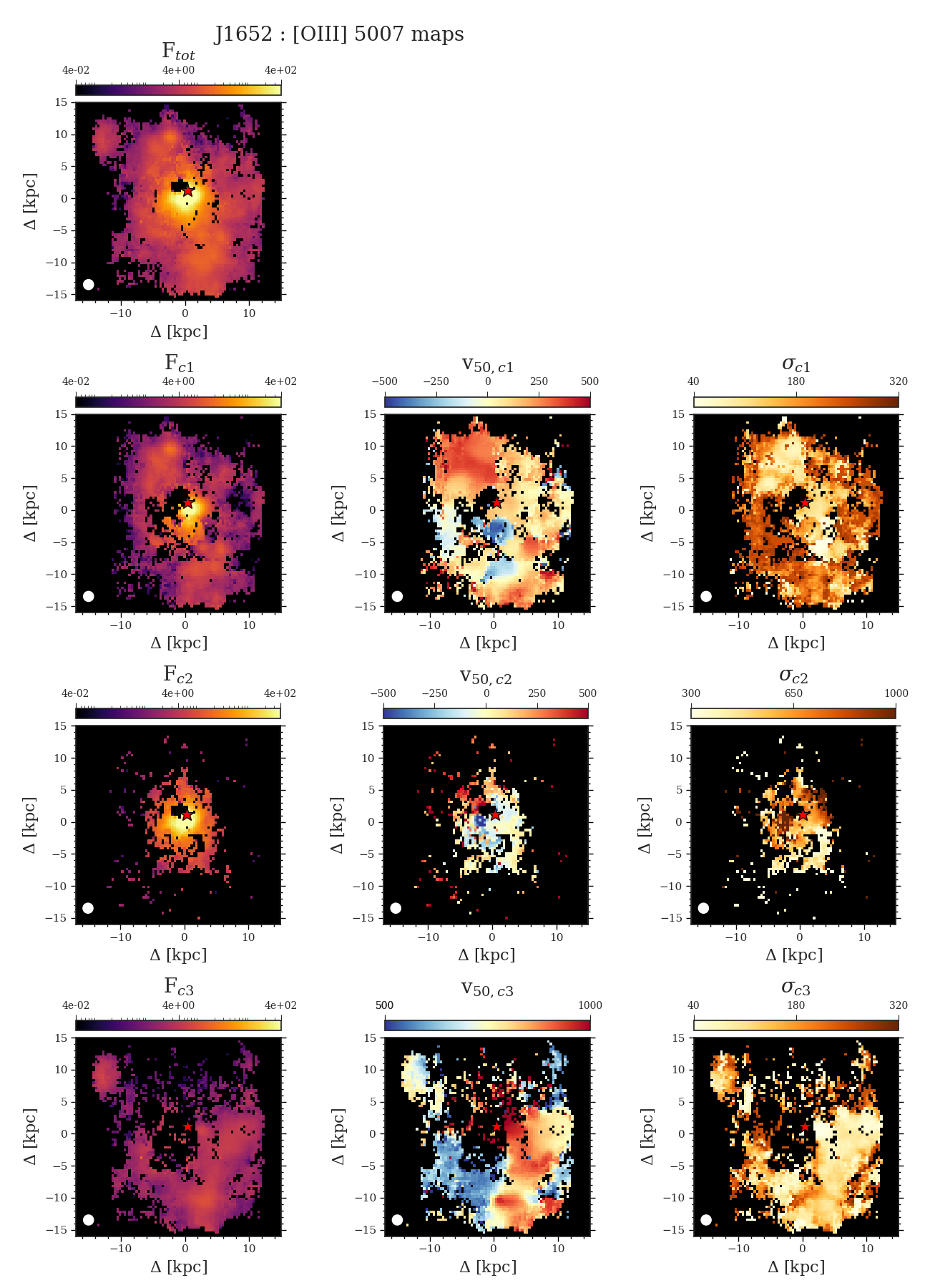}\\
    \caption{On the left, we present line-integrated \oiii\ 5007 \AA\ flux in units of $\times10^{-17}$\surff, in the middle, we present the radial velocity centroid ($v_{50}$) map in units of \kms\ and on the right the velocity dispersion maps for each detected kinematic components in J1652. In each map, ``c" stands for the Gaussian component number, where component 1 is associated with the systemic gas in the quasar host galaxy, component 2 with the outflow, and component 3 with the tidal tail feature and the neighboring galaxies. North is up, and east is to the left. The ellipse in the lower left corner shows the size of the NIRSpec IFU PSF.}
    \label{fig:components}
\end{figure*}

\begin{figure}
	\centering
	\includegraphics[width=1.0\linewidth]{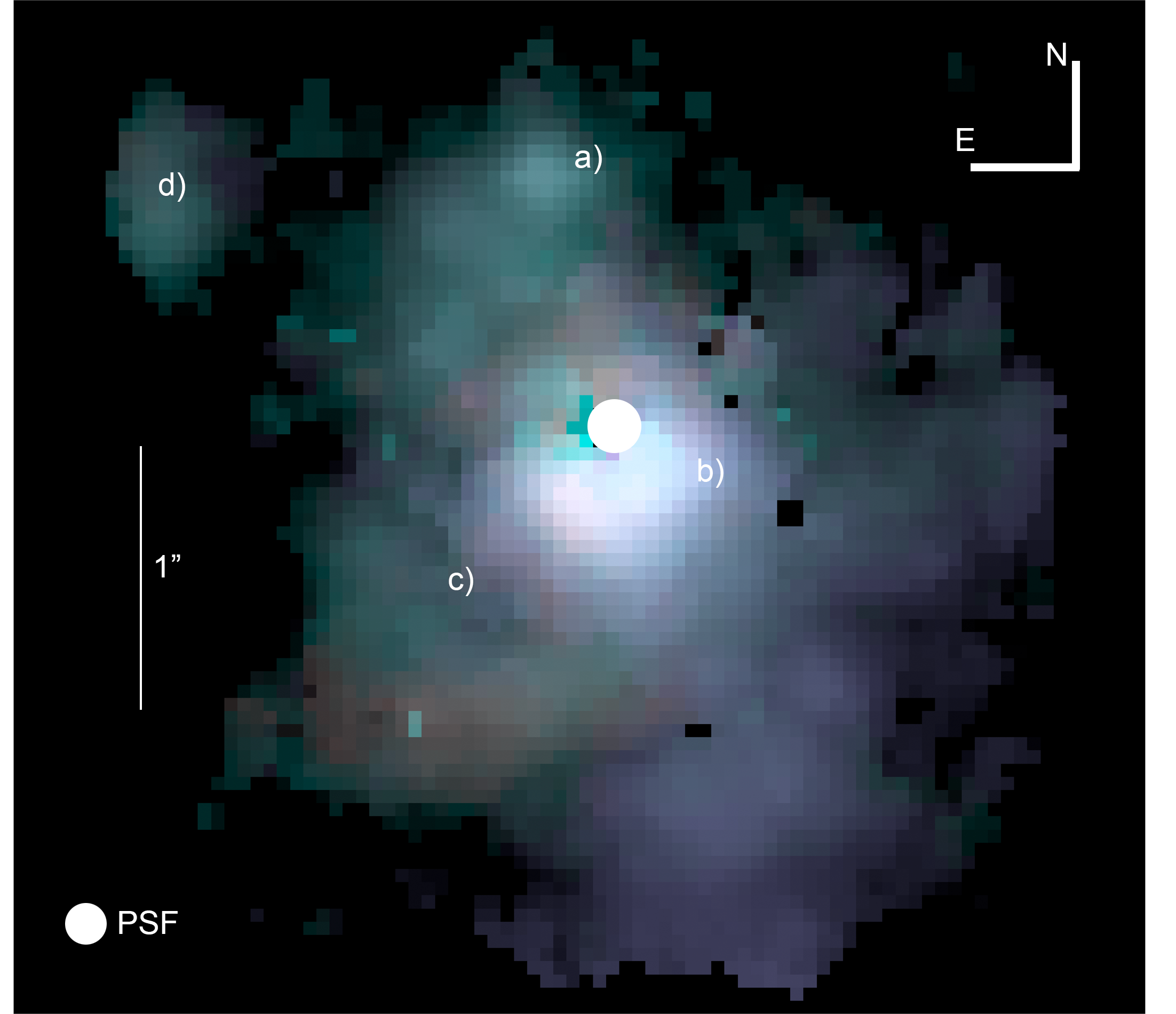}
	\caption{A color composite of ionized gas emission in the J1652 quasar host galaxy and the immediate environment. \oiii\ is assigned to blue, \ha\ is assigned to green, \nii\ is assigned to red and \sii\ is assigned to orange. Bar on the left represents 1 arcsecond, and the quasar is located behind the white ellipse near the center. The ellipse to the lower left represents the FWHM of the PSF. Regions for which spectra are extracted in Figure \ref{fig:decomposition} are labeled a-d. North is up, and east is to the left.}
	\label{fig:threecolor}
\end{figure}

\begin{figure}
	\centering
	\includegraphics[width=1.1\linewidth]{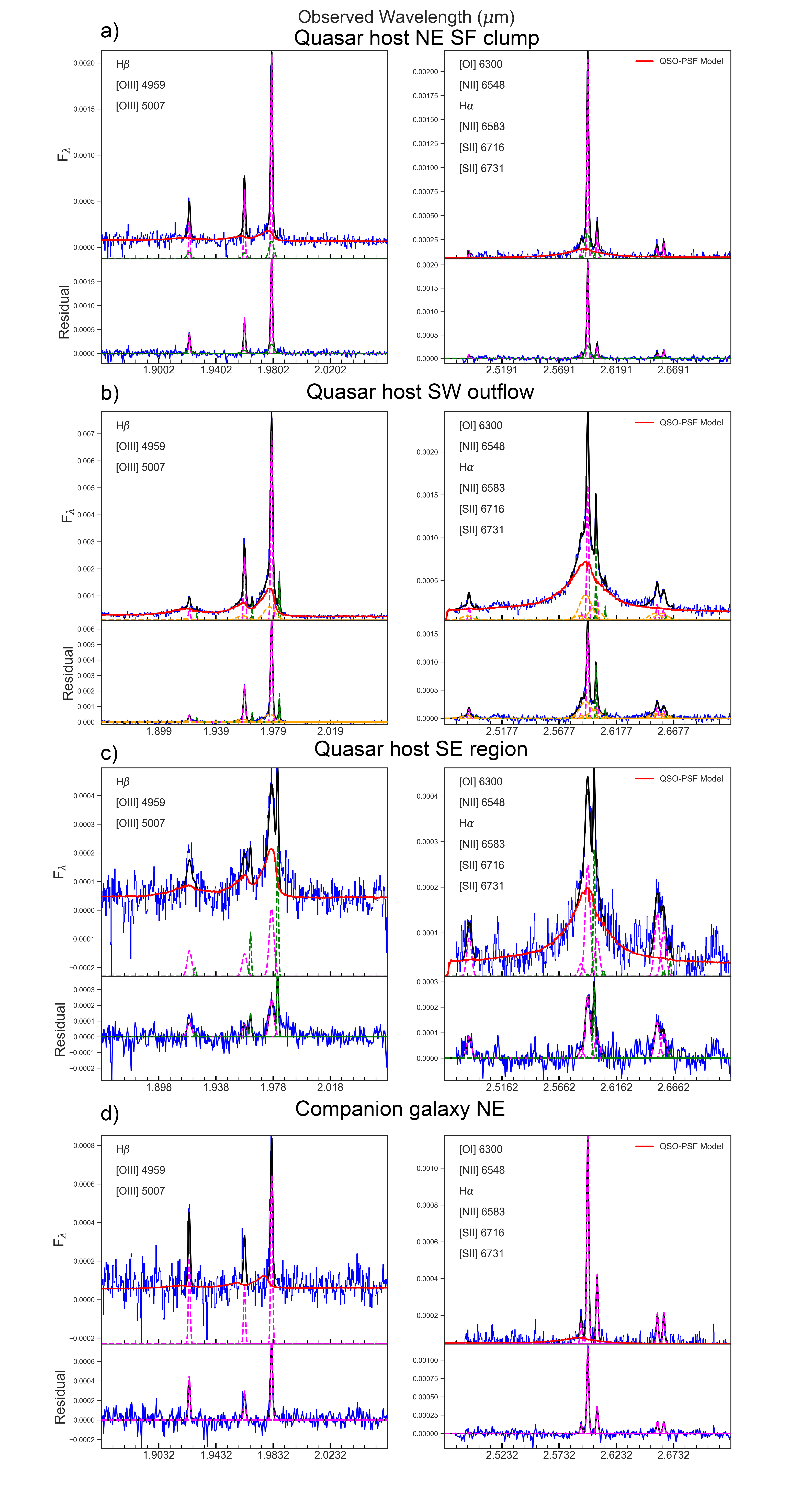}
	\caption{Examples of kinematic decomposition of the quasar host galaxy, outflow, and the nearby companions. We present multi-component Gaussian model fits to the \hb, \oiii, \oi, \ha, \nii, and \sii\ emission lines in a single spaxel of a region at systemic velocity in the quasar host galaxy, the outflow, a second region at systemic velocity orthogonal to the outflow and a companion galaxy towards the northeast. The top panel for each spectrum shows a multi-component Gaussian fit (dashed lines) to extended emission lines, while the black curve shows the total best fit. The red curve shows the spatially unresolved emission fit to that spaxel part of the quasar-PSF subtraction procedure in \qfit. The bottom panel shows the residuals after subtracting out the spatially unresolved emission, showcasing only the extended emission. The dashed curves show the individual multi-component Gaussian fit to the extended emission while.}
	\label{fig:decomposition}
\end{figure}

After PSF subtraction, we use \qfit\ to iterate over the spaxels in the entire NIRSpec field of view and fit each spectrum with a model consisting of a low-order polynomial continuum and emission lines. \qfit\ is initialized using the updated redshift of $z = 2.9489$ taken from \citet{Wylezalek22}. The fitting wavelength range corresponds to rest-frame wavelengths of 0.42-0.8 \micron, excluding the wavelength range of 0.598 - 0.627 \micron due to the NIRSpec detector gap.

We fit the entire cube with $n=1$, $n=2$, or $n=3$ Gaussian components for each emission line. The centroids and the velocity dispersions of each component are allowed to vary, but we assume that each Gaussian component has the same kinematic properties across all emission lines (i.e., they are `kinematically tied'; \citealt{zaka16b}), and we fit for intensities. The fitting uncertainties are those from \texttt{lmfit}, which is what performs $\chi^2$ minimization in \texttt{q3dfit}, and therefore the errors reflect the shape of the $\chi^2$ surfaces. We define a successful detection if the emission line peak in the data cube is $>3\sigma$, and the line width is greater than the instrumental width of the line-spread function. We detect H$\alpha$, H$\beta$, HeII $\lambda$4686, \oiii\ $\lambda$4959, 5007, [OI] $\lambda$6300, \nii\ $\lambda$6548, 6583, and [SII] $\lambda$6716, 6731. 

We find a minimum change in the $\chi^2$ value when increasing or decreasing the number of fitted Gaussian components that does not correspond to a worse or better fit. Hence we inspect each fit visually and decide to keep the fit with the least number of components that best fits the observed emission line profile.

\subsection{Component sorting and kinematics}

In many cases, it is undesirable and, in fact, incorrect to ascribe physical significance to individual Gaussian components in a multi-Gaussian fit. \citet{zaka14} demonstrated that in the spatially integrated spectra of powerful quasars, the lines were well-represented by a sum of multiple Gaussian components, but none of those components was in dynamical equilibrium with the host galaxy. In such cases, the rotation of the host galaxy and the blending of multiple velocities present within the outflow add up together in one integrated spectrum, and a sum of Gaussians is merely a way to quantify the profile in a noiseless manner rather than a physically motivated model of the galaxy and quasar outflow kinematics.

The situation is different in IFU spectroscopy. When observations are obtained at high spatial resolution, and when the same kinematic component can be traced across the field of view with minor variations in the velocity centroid and velocity dispersion, it becomes possible to ascribe the components to distinct dynamical and ionization mechanisms. 

Nonetheless, component sorting and identification with particular dynamical components of the system is non-trivial. After examining the velocity distribution, velocity dispersion distribution, and the morphology of all components in the multi-Gaussian component fits, we decided to categorize the components into three different classes. One is a narrow component with varying centroid velocity and velocity centroid near the systemic velocity of the quasar, so we define the `narrow systemic' emission arising in the gas with centroid velocity in the $[-500,500]$ \kms\ range with $\sigma_v<300$ \kms.  One is a broad component associated with the outflow (Vayner et al. 2023b - in prep.) with a characteristic velocity dispersion $\sigma_v>300$ \kms. And the last component is associated with narrow-line emission ($\sigma_v<300$ \kms) at very high-velocity offset ($|v|>500$ \kms) from the central galaxy and morphologically concentrated in clumps, so it is natural to interpret these features as companion galaxies, likely at a few tens of kpc and projected onto the NIRSpec field of view \citep{Wylezalek22}. For four of these companions, we also detect stellar emission \citep{zaka19}, confirming their nature as companion galaxies. The location of the companion galaxies is based on the location of the stellar continuum detected in NIRSpec observations combined with the location of ionized gas clumps and kinematics resembling local velocity gradients likely due to internal motion within the companion galaxies. Our velocity dispersion cut between the broad and narrow extended emission sufficiently and clearly divides the two kinematic components. The majority of the narrow emission shows velocity dispersion $\sigma_v<250$\kms, and the majority of the broad emission shows a velocity dispersion $\sigma_v>350$\kms, making clear bimodal distribution in the velocity dispersion in the SDSSJ1652 system. The broad emission on kpc scale is distinctly different from the spatially unresolved broad emission both in terms of velocity offset and dispersion. We find similarly extended emission using two different PSF subtraction methods \citep{Wylezalek22}, and the shape and velocity are consistent with ground-based AO observations \citep{vayner21a}. 

In Figure \ref{fig:components}, we show intensities, radial velocity offset, and dispersion of the \oiii\ emission line for the three kinematic components identified above, we mark the component associated with the systemic quasar host galaxy velocity as component (``c1"), gas associated with the outflow as ``c2" and emission associated with the tidal tail feature, and the neighboring galaxies as ``c3". In Figure \ref{fig:threecolor}, we show a multi-color composite consisting of the brightest detected lines (\oiii,\nii,\ha,\sii) across the entire NIRSpec field of view. In Figure \ref{fig:decomposition}, we show the emission lines along with the multi-Gaussian fits and PSF subtraction in the distinct kinematic regions (marked a-d) across the NIRSpec FOV with their spatial location marked in Figure \ref{fig:threecolor}. To summarize, thanks to the high spatial resolution of the data, we can separate the emission into three distinct kinematic components based on their velocity offsets and dispersions. Physically, one is associated with narrow emission in the quasar host galaxy, one with broad extended emission due to the quasar-driven outflow, and the last one with the narrow extended emission associated with the tidal tail and neighboring companion galaxies. In this paper, we focus on emission line ratios associated with these three distinct kinematic components. 

In \citet{zaka19}, we analyzed the F160W broad-band {\it HST} data for J1652 and concluded that the best-fitting two-dimensional Sersic profile had a Sersic index $n_s\simeq 3.4$, intermediate between disks ($n_s=1$) and ellipticals ($n_s>4$), although the quality of the fit was not very good due to strong residuals in modeling of the quasar PSF. The position angle of the long axis of the fit was $\simeq$26\deg\ East of North, and ellipticity of the fit was $\epsilon\simeq 0.3$, which is qualitatively similar to the orientation and shape of the stellar component seen in the {\it JWST} data (Ishikawa et al. in prep.). If the galaxy was a rotating disk, with its apparent ellipticity entirely due to projection effects, we would expect to see a velocity gradient perpendicular to the minor axis whose position angle at 116\deg\ East of North. We do not see a well-organized velocity field, and we, therefore, do not detect the host galaxy rotation and cannot identify its kinematic axis. 

\section{Discussion}
\label{sec:discussion}

Integral-field spectroscopy is a powerful tool to both spatially and spectrally map the extended emission around quasars and their host galaxies. NIRSpec on {\it JWST} enables unprecedented near-infrared integral-field capabilities in space for the first time, allowing detailed spectroscopic mapping of the rest-frame optical continuum and emission lines of high-redshift objects such as J1652 \citep{Wylezalek22}. In this Section, we use optical line diagnostics \citep{bald81, veil87} to investigate the ionization mechanisms producing line emission (Section \ref{sec:ionization}), physical conditions in different parts of the host galaxy and its environments (Section \ref{sec:physical}), and morphology of the different components (Section \ref{sec:morphology}). 

\subsection{Ionization mechanisms}
\label{sec:ionization}

\begin{figure*}
	\centering
        \includegraphics[width=1\linewidth]{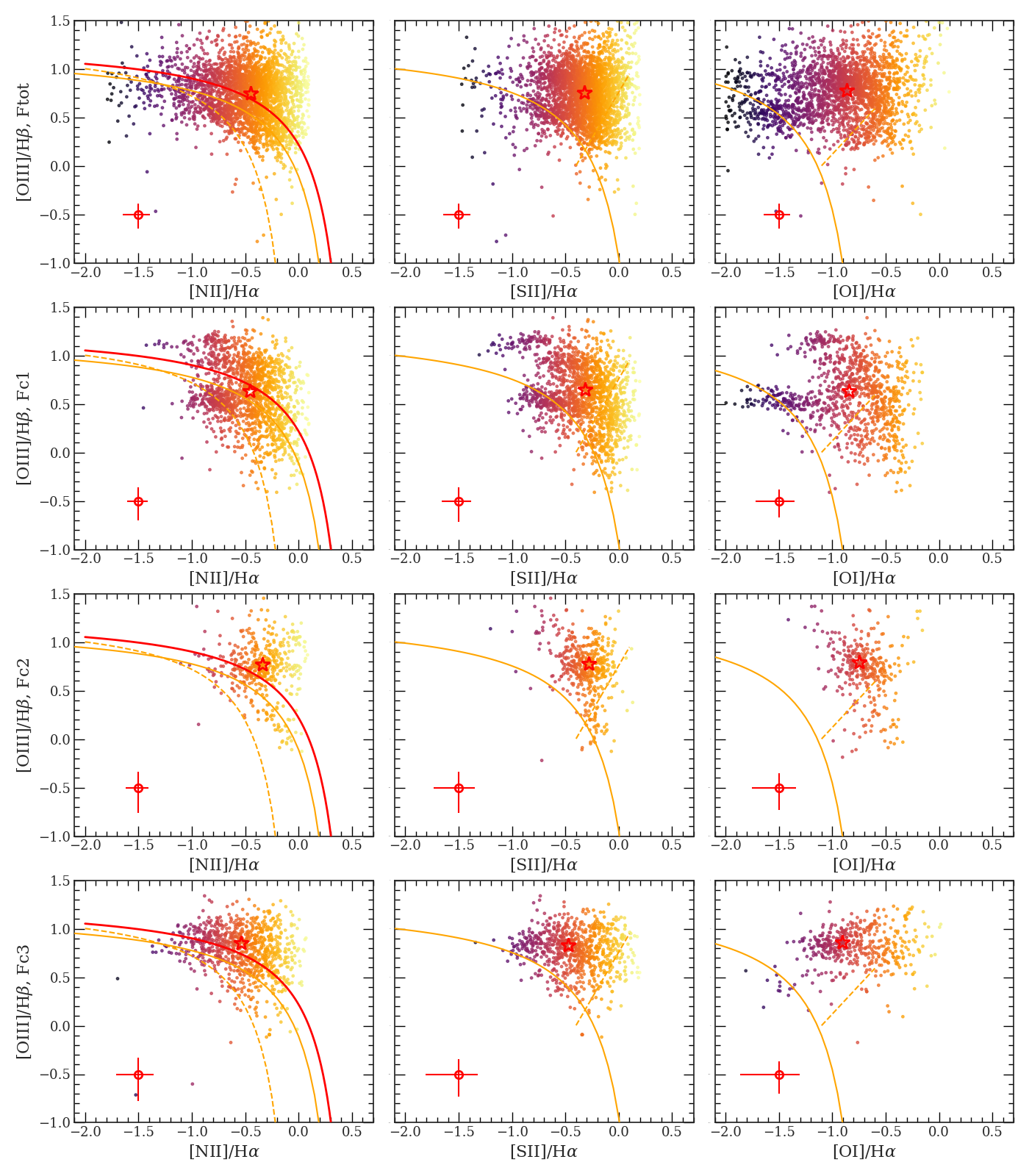}
	\caption{Spatially-resolved emission-line diagnostic diagrams for J1652. The top row shows the line ratios for the kinematically integrated line ratios. The bottom three rows are for the three distinct kinematic components. The star represents the average emission line flux ratio for the individual kinematic component. The circle point to the lower left shows the median error on the emission line ratios. The points are color-coded to match the maps presented in Figure \ref{fig:BPTmaps}. The dashed orange line shows the location of the star formation abundance sequence at $z=0$ from \citet{kauf03a} on the \oiiihb\ vs. \niiha\ line ratio diagram and the separation between Seyferts and LINERs on the \oiiihb\ vs. \siiha\ and \oiha\ diagrams, and the solid orange line shows the delineation between star formation and AGN photoionization based on theoretical models by \citet{kewl01} for low redshift, while the solid red line shows the upper bound on the theoretical abundance sequence with evolving gas conditions as a function of redshift at $z=3$ \citep{Kewley13b}.}
	\label{fig:bpt}
\end{figure*}

\begin{figure*}
	\centering
	\includegraphics[width=1\linewidth]{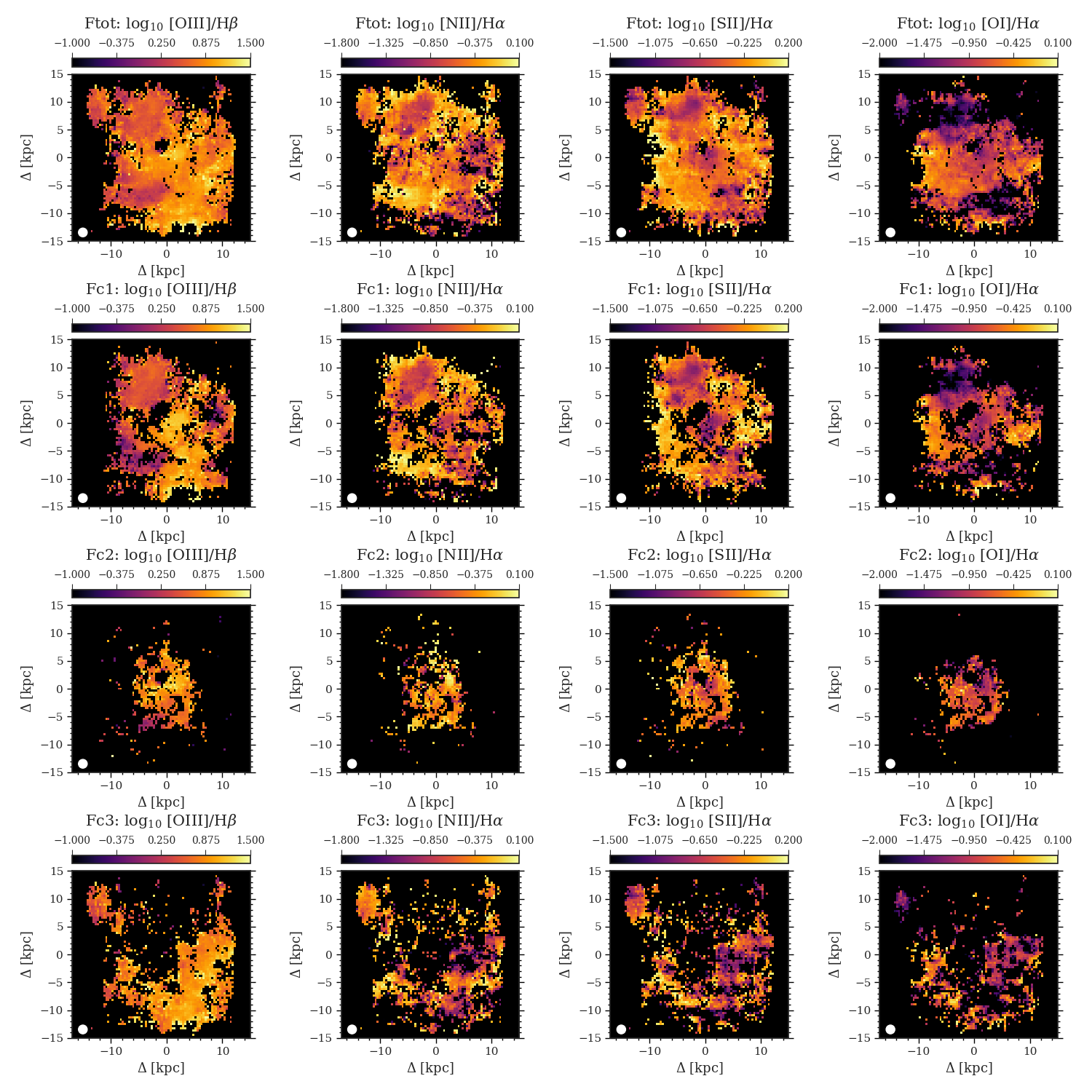}
	\caption{Maps of emission line ratios. We present emission line ratio maps for \oiiihb, \niiha, \siiha, and \oiha\ for each distinct kinematic component in the bottom three rows. The top row shows the emission line ratios integrated over the three kinematic components. The ellipse in the lower left corner shows the size of the NIRSpec IFU PSF. North is up, and east is to the left.}
	\label{fig:BPTmaps}
\end{figure*}

We measure \oiii/\hb, \sii/\ha, \oi/\ha, and \nii/\ha\ line ratios for each kinematic component. These measurements are shown in Figure \ref{fig:bpt} for both integrated line ratios and for each component independently, together with the theoretical lines delineating photoionization by star formation, photoionization by a quasar, and shock ionization. In Figure \ref{fig:BPTmaps} we show the 2D maps associated with each of the emission line ratios for each kinematic component with color scaling to match Figure \ref{fig:bpt}. 

\subsubsection*{Quasar ionization cone:}
Gas in the kinematic component associated with the quasar host galaxy (Component 1 in Figures \ref{fig:bpt},\ref{fig:BPTmaps}) shows a bimodal distribution in all line ratio diagrams. We see elevated \oiiihb\ line ratios towards the southwest, the same direction in which the photons must propagate from the nucleus and scatter off the material in the interstellar medium to explain the available spectropolarimetric observations \citep{alex18} and the extended UV continuum resolved in {\it HST} observations (Vayner et al. 2023b - in prep.). We classify this region as the ionization cone due to the quasar emission. 

Within this region, we also see several spatially clumpy regions that are associated with low \niiha\ line ratios, likely indicating that star formation within these kpc-scale clumps might be partially contributing to the photoionization of the gas. However, the \oiiihb\ ratio in these clumpy regions is too high for star formation to be a dominant source of ionization.

\subsubsection*{Quasar host galaxy star-forming region:}

In the kinematic component associated with narrow gas near systemic velocity (component 1), towards the northeast, we detect three distinct kpc-scale clumps that show low \niiha\ ($<-0.5$) and relatively low \oiiihb\ ($<0.5$) emission line ratios. On the classic $z=0$ line ratio diagram, these clumps fall above the star-forming sequence and are close to the transition zone between star formation and AGN photoionization. 

However, at higher redshifts, a lot of star-forming galaxies are above the $z=0$ star-forming sequence \citep{Strom17,Strom18} due to harder ionization spectra, lower gas and stellar [Fe/H] metallicities \citep{Strom17,Strom18,Sanders23} and overall denser interstellar medium (ISM; \citealt{Sanders15}). The red curve in Figure \ref{fig:bpt} shows the delineation between star formation and AGN photoionization from \citet{Kewley13b}, which takes into account these redshift-dependent changes in typical physical conditions. Using these criteria, the J1652 clumps in the northeast fall within the maximum allowed line ratios produced by star formation. Hence, star formation is likely the primary source of ionization in these three clumps. Furthermore, this photoionization model best explains the delineation of the bimodal distribution in the line ratios that we see in this kinematic component. However, using the \niiha\ and \oiiihb\ ratios alone, we found a few spaxels left over that contain relatively high \oiha\ and \siiha\ emission line ratios, inconsistent with star formation. Hence, in addition to the \niiha\ and \oiiihb\ line ratios, we require that the line ratios fall within the star-forming portion of the \oiiihb\ vs. \siiha\ diagram. The reason for selecting the \siiha\ line ratios over \oiha\ is due to a larger coverage of the \sii\ in the NIRSpec FOV.

\subsubsection*{Quasar host galaxy outflow region ionization mechanisms:}
The spatially-extended, kinematically-broad emission associated with the quasar-driven outflow (`Fc2') discussed by Vayner et al. (2022b - in prep.) and also detected in the ground-based data \citep{vayner21a} -- more prominent toward the southwestern direction from the nucleus -- is primarily photoionized by the quasar. Spatially, the entirety of this component lies within the quasar photoionization cone identified via the line ratios of the narrow quasar host galaxy component (`Fc1').

\subsubsection*{Companion galaxies and tidal tail ionization:}
Component 3 (`Fc3') associated with the tidal tail emission and nearby satellite galaxies selected based on emitting gas with a narrow velocity dispersion and a large ($V>500$\kms) velocity offset relative to systemic, shows a wide range of emission-line ratios. A large portion of the region in the southwestern direction of the quasar photoionization cone shows clear evidence for elevated \oiiihb\ emission line ratios, consistent with quasar photoionization. Galaxies towards the west and southwest have their ISM partially photoionized by the quasar, indicating evidence for the impact of gas heating by the quasar on the immediate satellite galaxy population. The galaxies to the west and southwest also show some spaxels consistent with star formation photoionization. The galaxy towards the northeast, which falls outside the quasar photoionization cone, shows evidence for star formation ionization with emission line ratios similar to the clumpy regions seen in the quasar host galaxy. The tidal tail feature connecting the galaxies is primarily photoionized by the quasar.

\subsubsection*{Quasar host galaxy shocks region:}
\begin{figure*}
    \centering
    \includegraphics[width=0.7\linewidth]{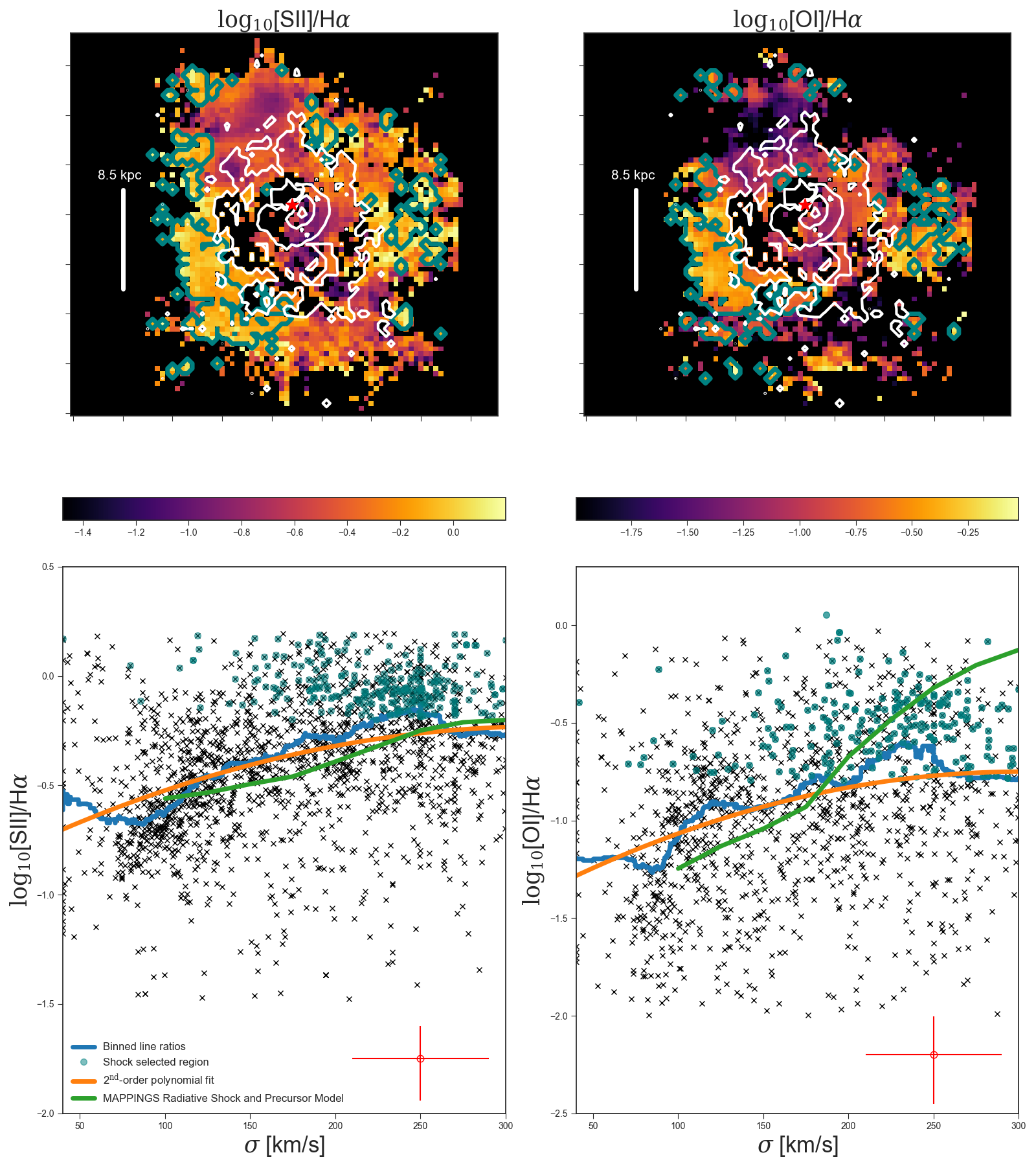}
    \caption{Shock diagnostics in the kinematic component of the J1652 quasar host galaxy. The top panels show the \siiha\, and \oiha\ emission line ratio maps where the bar represents 8.5 kpc and the red star shows the location of the quasar. Teal contours show the shock ionization regions orthogonal to the outflow shown in white contours. We show the emission line ratios of \siiha\ and \oiha\ as a function of the emission line dispersion in the bottom panel with the black points. The red point in the lower right corner shows the median emission line ratio and velocity dispersion errors. We observe, on average enhanced line ratios with higher velocity dispersion in regions perpendicular to the quasar outflow and the ionization cone shown in the green points. Binned line ratios are shown with the blue curves. The orange curves represent a 2nd-degree polynomial fit to all data points. The green curve shows a radiative shock model from \cite{alle08} for an electron density of 100 \eden, solar metallicity, and magnetic field strength of 10 $\mu$G cm$^{3/2}$.}
    \label{fig:ratio_dispersion}
\end{figure*}

In the quasar host galaxy kinematic component, the \oi/\ha\ map and (to a lesser degree) the \sii/\ha\ map show enhanced values in the East-West direction, roughly orthogonal to the direction of the photoionized cones. Enhanced \oi/\ha\ values appear to be specifically anti-correlated with \oiii/\hb\ values that systematically show lower (\logohb $< 0.5$) values. We further see a general trend where on average, there are higher \oiha\ and \siiha\ line ratios at higher velocity dispersion within the narrow kinematic component associated with the quasar host galaxy (Figure \ref{fig:ratio_dispersion}), suggesting shock ionization that extends perpendicular to the quasar outflow, and quasar ionization cone \citep{rich15}. Furthermore, compared to the emission line ratios predicted by radiative shock models, we find that the points at $\sigma>150$ \kms\ and orthogonal to the outflow region are consistent with these models \citep{alle08}. We find a radiative shock model with an ISM electron density of 100 \eden, solar metallicity, and magnetic parameter (B/$\sqrt{n}$) of 10 $\mu$G cm$^{3/2}$ best fits the observed line ratios orthogonal to the outflow (Figure \ref{fig:ratio_dispersion}). We find that higher electron density models under-predict the observed line ratios at all velocities. The large scatter that we see in Figure \ref{fig:ratio_dispersion} likely indicates that there are still other photoionization mechanisms at play. The line ratios in the shock-ionization part of the diagnostic diagrams can have many different origins. In some objects, these lines originate in shocks produced either by jets \citep{ogle10,Ogle12,Lanz15}, galactic winds \citep{Veilleux95} or by galactic collisions \citep{Appleton06,rich11, rich15}, or sometimes unusual ionization sources masquerade as shock ionization \citep{belf16}. 

In our case, we know that there is a powerful quasar-driven outflow. The propagation of galactic outflows is strongly affected by the distribution and clumping of the interstellar medium. Winds can hydrodynamically curve around obstacles \citep{wagn13} and reach the parts of the galaxy that are not directly illuminated by the quasar within the photo-ionization cone, which is shaped by circumnuclear obscuration. The direct signatures of quasar winds impacting the host galaxy by driving turbulent shocks have been long sought, with the majority of the evidence in lower luminosity AGN in very nearby galaxies \citep{Veilleux95,Allen99} and in nearby quasars through the enhancement of certain shock diagnostics as a function of wind kinematics seen in integrated quasar spectra \citep{zaka14}. More recently, with the advent of integral field spectroscopy, direct evidence for shocks visible just outside of the quasar illumination cone has started to accumulate. \citet{riff21} found direct evidence for shock ionization in Mrk79 and Mrk348 that is orthogonal to the ionization axis, in which shocks can be more easily observed in regions that are shielded from the quasar's radiation field by the dusty nuclear torus. Similarly, \citet{Leung21} found a region of low-ionization, high-dispersion gas in Mrk273, which is neatly orthogonal to the photo-ionization cone. The implication is that the wind opening angle may be significantly larger than the opening angle of the ionization cone, where the gas is most visible. Here we present the first direct evidence for this phenomenon outside of the local universe, in a high-redshift quasar.  

J1652 has shocks extended on the galactic scales, yet, it is difficult to confidently determine whether the shocks are driven by the quasar outflow or by the potential merger-induced inflows. J1652 is likely experiencing a major merger, with an extended tidal tail seen in the {\it HST} data \citep{zaka19} and multiple companions within several tens of kpc from the host \citep{Wylezalek22}. Unfortunately, the small field of view of NIRSpec prevents us from conclusively determining whether there is any shocked emission specifically associated with the tidal tail. Both the merger and the outflow could be contributing to the shocked emission. However, there is a hint of increased \oiha\ and \siiha\ line ratios where there are lower \oiiihb\ emission line ratios towards the edge of the outflow as seen in the third panel of Figure \ref{fig:BPTmaps} in the \siiha\ and \oiha\ line ratio maps, indicating that the gas in the outflow transitions from being predominantly photoionized by the quasar to shocks ionization due to interactions of the outflow with the ISM of the host galaxy. This suggests that the outflow may be the dominant source of shock heating in both the narrow emission and the outflow region, providing hints for quasar feedback through the shock heating of the ISM. Over the shock region, we don't see any evidence per spaxel of broad emission with velocity widths similar to that found in the individual spaxels over the outflow. The more diffuse nature of the outflow on larger scales, combined with the loss of quasar photoionization outside the ionization cone, makes it far more difficult to detect the broader emission from the outflow. Narrower turbulent emission that is likely a consequence of the outflow is easier to detect due to high surface brightness.


\subsection{Physical conditions}
\label{sec:physical}

The unprecedented sensitivity of {\it JWST} allows us for the first time to measure the \sii\ 6717 \AA\ \& 6731 \AA\ emission-line ratios, commonly used as electron density diagnostics, in distinct regions of distant galaxies. The \sii\ lines are sensitive to ionized gas density for temperatures $\sim10^{4-5}$K and densities between 100-10,000 \eden. We can measure the emission-line ratio at high significance ($>3\sigma$) in the northeast star-forming clump region in the quasar host galaxy and in the merging galaxy towards the northeast. Over the outflow region, measuring the \sii\ emission line ratio is more challenging due to the larger kinematic widths and, therefore, more significant blending. However, we find a decrease in the velocity dispersion with radius and can adequately fit both of the emission lines. We use the \texttt{getTempDen} routine from the \texttt{PyNeb} \citep{Luridiana15} package to derive the electron density using a gas temperature of 10,000 K. In the star-forming clumps of the quasar host, we measure an average electron density of 630$_{-300}^{+1400}$ \eden\ with a few regions getting to densities as high as 3,000-4,000 \eden\ (Figure \ref{fig:haSFR}). In the outflow region, we measure a median value of 870 $^{+2000}_{-700}$ \eden, and in the neighboring galaxy towards the northeast, a median value of 630 $^{+2000}_{-470}$ \eden. 


In addition to measuring the electron density, we can measure the metallicity of the gas in the star-forming clumps. We do not detect the \oiii\ 4363 \AA\ emission line across the star-forming region, preventing us from using a direct approach to measuring the gas-phase metallicity. For the quasar host galaxy, neighboring galaxies, and tidal tail, we isolate star-forming regions based on the diagnostic line ratios as discussed in Section \ref{sec:ionization}. We use the \niiha (N2) emission line ratio and the oxygen to hydrogen abundance empirical calibration from \cite{Marino13}. We find \oh\ values of 8.29-8.5 in the quasar host galaxy and 8.3-8.5 in the neighboring galaxies and tidal tail feature. We measure similar metallicities in each spaxel of the star-forming regions using the empirical oxygen to hydrogen abundances vs. log(\oiii/\hb$\times$\ha/\nii) (O3N2) calibration. We notice a minor positive gradient in the metallicity as a function of distance from the star-forming clump centers in the quasar host galaxy, most noticeably centered on the brightest northeast clump in Figure \ref{fig:haSFR}, likely indicating a recent accretion of new lower metallicity gas that triggered the star formation in the northeast clumps \citep{Queyrel12}. The gradient is present in both \oh\ abundance maps measured using the N2 and O3N2 calibration methods. Positive metallicity gradients are commonly found in star-forming galaxies that have recently undergone a merger. The ongoing merger in the J1652 system may be a partial cause of the observed gradient in the star-forming region \citep{Queyrel12,Jones13}. Outflows driven by stellar feedback may be expelling the higher metallicity gas towards larger radii away from the clumps, additionally helping cause the observed positive gradient \citep{Wang19}. Indeed we find evidence for outflows in these clumps due to broader blueshifted nebular emission lines relative to the narrow emission in the star forming clumps (Figure \ref{fig:decomposition}), similar to what has been found in clumps of star-forming galaxies at $z\sim2$ \citep{Genzel11}. 

Observations of star-forming galaxies at an increasingly higher spatial resolution near cosmic noon often find that the star-forming clumps resolve into smaller substructures, indicating that these large kpc scale clumps often contain substructures composed of several HII regions \citep{Cosens18,Cava18,Claeyssens23}. Given the 200 mas resolution of our observations, it is very likely that each star-forming clump is a composition of many smaller subclumps; hence the gradient in \niiha\ emission line ratio is likely a result of metallicity changes as a function of radius rather than changes in gas ionization as a function of the radius of a single HII region.

We find elevated ionized gas densities, lower metallicities, and emission line ratios consistent with what is found in star-forming galaxies at $z\sim2.3$ \citep{Sanders15,Strom17,Strom18}, further supporting our use of theoretical models that involve lower metallicity gas, harsher ionization and denser ISM \citep{Kewley13b} than those in local galaxies at the same stellar mass \citep{kauf03a}. Photoionization from the quasar of low metallicity gas is unlikely to cause ionization of the clumps towards the northeast since their \oiiihb\ values are lower than what is expected from quasar photoionization of low-metallicity (\oh$<$8.3) gas \citep{Kewley13b}.

We find elevated \ha/\hb\ emission line ratios ($>2.8$) in the star-forming clumps of the quasar host galaxy, the outflow region, and in the companion galaxy towards the northeast. These ratios are above what is expected for typical gas-phase conditions in the ISM of galaxies and radiation produced by recombination \citep{oste06}. This likely indicates the presence of dust within these regions that is causing reddening. Using the \citet{Calzetti00} extinction law and assuming Case B recombination for the intrinsic \ha/\hb\ emission line ratio, we find $V$-band extinction of 0.5$-$3 mag in the star-forming clumps towards the northeast in the quasar host galaxy, in the outflow, and in the companion galaxy towards the northeast. For the northeast star-forming clumps, we see an increase in the level of extinction away from the brightest parts of the clumps. This increase disfavors the scenario that there is less shielding from quasar photoionization in the more flux diffuse regions, and the entirety of our selected region is consistent with star-formation photoionization. Comparing all kinematic regions over the entire source, we find that the spaxels most consistent with star formation have the highest extinction values. V-band extinction magnitudes of 1-3 in the northeast star-forming region likely indicate sufficient self-shielding from the quasar UV ionization given that the UV optical depth due to dust absorption ($\tau_{UV}$) is $>$ 6-20 assuming the \cite{Calzetti00} extinction law. The expected hydrogen column densities in the regions with measured dust extinction, assuming a Milky-Way dust extinction, are 2-6 $\times10^{21}$ cm$^{-2}$ \citep{Guver09}.

\subsection{Morphology of the ionized gas in the host}
\label{sec:morphology}

In this section, we describe the morphology and extent of the kinematically-narrow emission associated with the quasar host galaxy -- within 500 \kms\ from the systemic velocity. A portion of this kinematic component is consistent with being photo-ionized by star formation, and the integrated flux distribution is shown in Figure \ref{fig:components}, second row. We see an overall orientation and ellipticity of the emission, which is consistent with those of the broad-band emission likely due to stellar light seen in the {\it HST} images \citep{zaka19}. 

The high-resolution {\it JWST} observations reveal that the line emission is very clumpy, with clumps of sizes 2-3 kpc and H$\alpha$ luminosities of $1-10\times10^{41}$ \ergs. We find at least nine individual clumps. We then isolate the spaxels in these clumps that fall within the star formation region of the line-ratio diagram as outlined by the red line in Figure \ref{fig:bpt}, neglecting \ha\ emission due to quasar photoionization. We find a total \ha\ flux associated with star formation of $1\times10^{-16}$ \ferg, which translates to a star formation rate of $69$ \myr\ using the \citet{kenn98} empirically derived conversion from \ha\ luminosity to a star formation rate. After correcting for extinction, these values increase to $3\times10^{-16}$ \ferg and $200$ \myr.

\begin{figure*}
    \centering
    \includegraphics[width=1\textwidth]{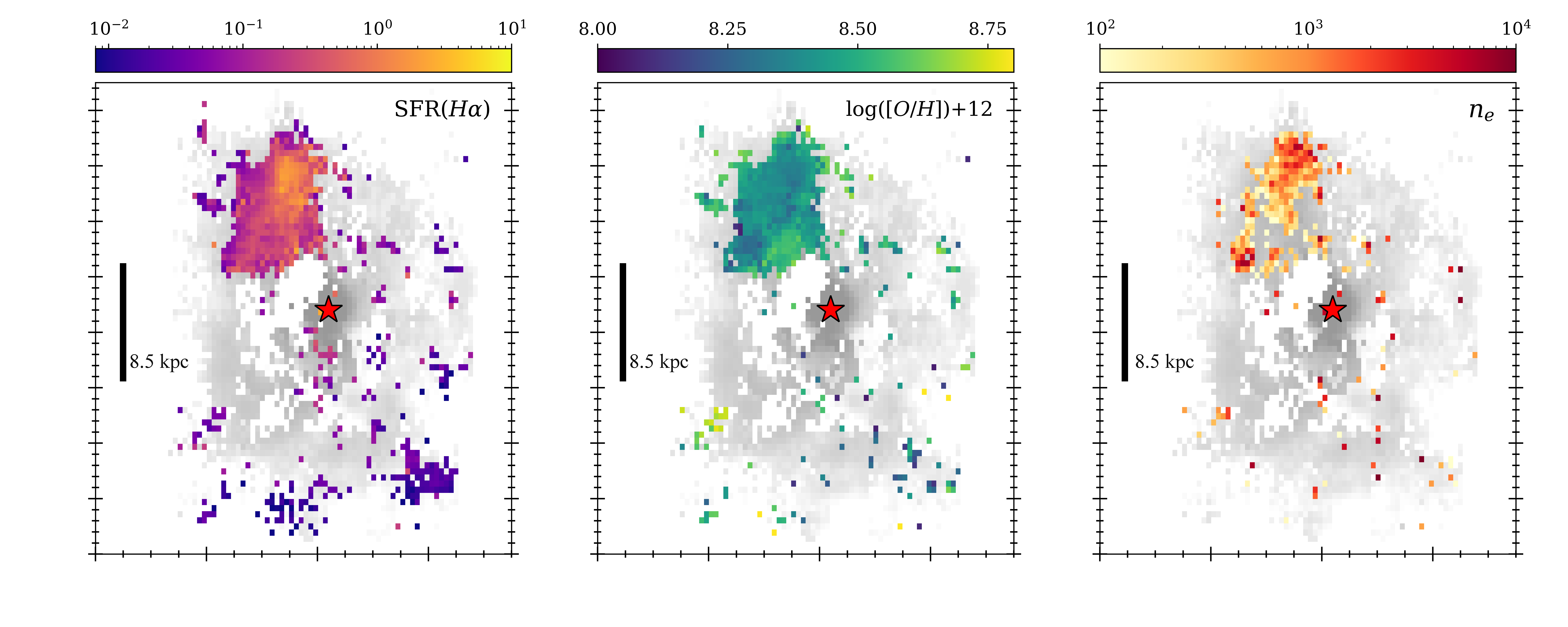}
    \\
    \caption{On the left, we present the extinction-corrected, \ha-derived star-formation rate map in the quasar host galaxy of J1652, only selecting spaxels with emission line ratios consistent with star formation ionization. In the center, we show the metallicity map. On the right, we show the electron density map. In each plot, we show the total \ha\ map in gray. We observe the brightest star-forming region to have lower metallicity and elevated electron density.}
    \label{fig:haSFR}
\end{figure*}

In Figure \ref{fig:haSFR}, we show the star-formation rate map based on the extinction corrected \ha\ emission. We can see that the region ionized exclusively by star formation, with no contribution from the quasar, is concentrated toward the northeast of the central quasar. This star formation rate is likely a lower limit as we are missing the obscured star-forming regions, as well as star-forming regions within quasar ionization cones where gas is `fried' by quasar radiation and excluded from our calculations by the line-ratio diagrams. 

The primary quasar photoionization cone and kinematically broad outflow are both directed to the south-west of the nucleus \citep{vayner21a}, but a faint north-eastern counter-cone with a much smaller physical extent is also detected via its kinematics \citep{Wylezalek22}. J1652 is an extremely powerful quasar, with quasar photoionization being the dominant ionization process across most of the observed emission. In order for the ionized gas to display star-formation-like line ratios in the northeastern quadrant, it must either be protected from quasar emission or star formation needs to dominate the overall ionization in the surrounding gas clouds. We hypothesize that the star-forming region identified in Figure \ref{fig:haSFR}, left, is physically in front of the quasar counter-cone and outside of the quasar photo-ionization region. This is suggested both by the faintness of the counter-cone, which in this scenario may be obscured by the foreground parts of the J1652 host galaxy hosting star-forming clumps (this region reaches $\rm A_V=3$ mag), and by the kinematics of the counter-cone -- redshifted, i.e., directed away from the observer \citep{Wylezalek22}. 

Although all star formation is ultimately regulated by inflows and outflows of gas, high-redshift star formation is dominated by somewhat different processes and is somewhat morphologically distinct from star formation in the local universe \citep{deke09b, Claeyssens23}. Most importantly, in the phenomenon known as `cosmic downsizing' \citep{hall18}, star formation at high redshift predominantly occurs in the most massive halos through high-rate accretion by gas streams and gas-rich mergers, whereas in the local universe, the most massive halos are inhabited by passively evolving, gas-poor galaxies. J1652, which is hosted by a massive galaxy \citep{zaka19} within a likely protocluster \citep{Wylezalek22}, is a prime example of this phenomenon. 

Morphologically, high-redshift star formation is significantly clumpier \citep{guo15} and more dynamically hot than that at low redshifts -- specifically, star formation rarely proceeds in well-organized star-forming disks, and in rare disk examples, the typical velocity dispersion of the gas is much higher than that in the local ones \citep{law12}. This is also in line with our observations of J1652, where multiple gas clumps are seen with a wide range of velocities, where the morphology of the stellar component -- disk vs. spheroid -- remains ambiguous, and where, despite the high galaxy mass, we have not been able to detect a systematic rotation signature. 

The giant star-forming clumps do not necessarily require mergers and, in simulations, can be produced as a result of secular processes such as cold stream accretion \citep{deke09b} with subsequent clump mergers \citep{tamb15}. All of these processes can be in play in J1652, but in addition, its tidal tail indicates a possibility of a recent major merger. Statistically, major mergers have been linked to brighter and larger clumps in high-redshift star-forming galaxies \citep{ribe17,Genzel11}. 

\begin{figure*}
	\centering
\includegraphics[width=0.9\linewidth,trim={1.5cm 1.5cm 3cm 1.5cm},clip]{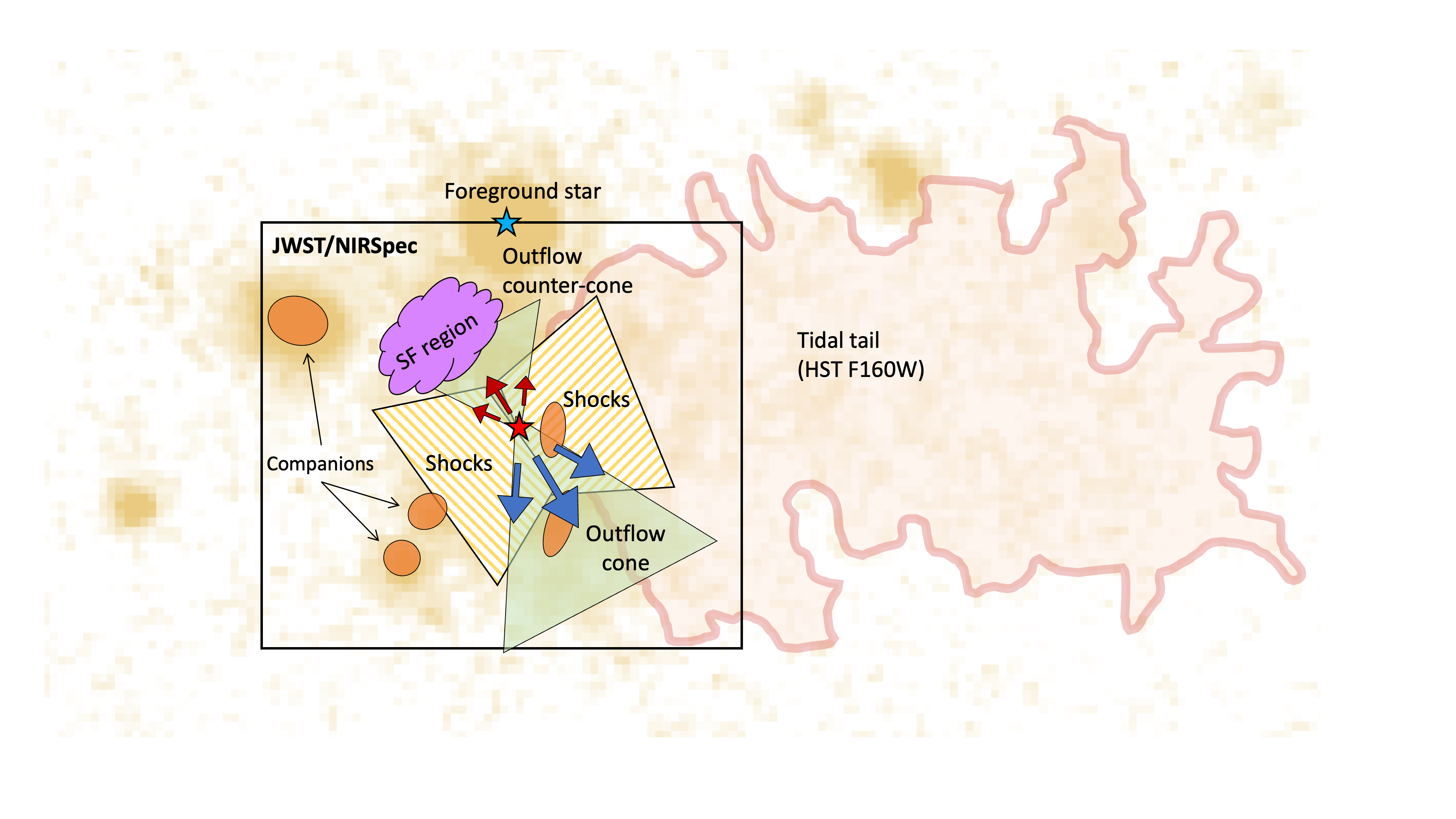}
	\caption{The schematic of the major components of J1652 as seen in the {\it JWST} and in the {\it HST} data, with the {\it HST} WFC3 data (F160W, corresponding to rest-frame $B$-band; \citealt{zaka19}) used as the background image. The gray outline roughly traces the \oiii\ emission shown in \cite{Wylezalek22}.}
	\label{fig:schematic}
\end{figure*}

\section{Conclusions}
\label{sec:conclusions}

In this paper, we present {\it JWST} NIRSpec integral-field unit observations of J1652, a powerful red quasar at redshift at $z\sim3$. J1652 belongs to the population of extremely red quasars, which are selected from optical and infrared surveys based on their colors and rest-frame ultraviolet emission line properties \citep{ross15, hama17}. Amongst quasars of all types and all luminosities, they are unique in exhibiting extremely fast winds seen in forbidden lines of warm ionized gas \citep{zaka16b, perr19}. We analyze the observations using \qfit, which subtracts the PSF from the data cube and fits continuum and emission-line models to the residual host galaxy data. In our previous papers based on this dataset we demonstrated that J1652 is in a very active environment, with multiple companion galaxies within the small field of view of NIRSpec moving with high velocities \citep{Wylezalek22}, and we confirmed the presence of an extremely powerful galactic-scale quasar-driven wind whose energetics and physical conditions can be measured with {\it JWST} for the first time (Vayner et al. 2023b - in prep.). 

In this paper, we present the analysis of the ionization physics of the gas based on the emission-line diagnostic diagrams for different kinematic components. We separate the Gaussian components from the multi-Gaussian emission-line fits into broad and likely due to the quasar-driven outflow and narrow, which may be due to companion galaxies or to clumps of star formation in the galaxy itself. We confirm at least nine clumps within 500 \kms\ of the systemic velocity characterized by a narrow velocity dispersion. The morphology is too disturbed and broken into clumps for us to see any evidence of an organized velocity field due to galaxy rotation. This, combined with a Sersic index of $n_s=3.4$ measured from the {\it HST} data \citep{zaka19}, which is intermediate between disks and ellipticals, prevents us from conclusively determining the dynamical state of the host galaxy. 

Standard emission-line diagnostics \citep{bald81, veil87}, which can be measured for integrated lines or for individual kinematic components, reveal a complex superposition of regions dominated by photoionization by the quasar, by star formation and by shocks. Photoionization by the quasar is primarily in the southwest direction, along the same axis as the one expected from the ground-based spectropolarimetric observations \citep{alex18}, which probe circumnuclear geometry of obscuration. The brighter southwestern ionization cone, blue-shifted and likely pointing somewhat toward the observer, coincides with the location of the rest-frame UV extended emission likely due to scattered light \citep{vayner21a}. We also detect the much weaker redshifted counter-cone on the opposite northeastern side of the nucleus. Therefore, the standard geometric unification model developed for nearby low-power active nuclei \citep{anto93} applies to our high-redshift, extremely powerful obscured quasar as well. 

We clearly detect shock-ionized emission in the regions perpendicular to the main quasar ionization cone. This component is seen in both \oi/H$\alpha$ and \sii/H$\alpha$ maps. The shocks may be due to collisions with companion galaxies -- or, more likely, to the quasar-driven wind. It has long been suggested by numerical simulations that quasar-driven winds can hydrodynamically go around obstacles and shock-excite the gas outside of the ionization cone of the quasar \citep{wagn13}. If the moving wind is also illuminated by the quasar, then in the clouds affected by both shocks and photo-ionization, the latter dominates, and the shock signatures are very hard to detect. However, the gas outside of the ionization cone is protected from quasar emission, and the weaker shock signatures become visible. Our detection of the off-axis shocked emission strongly suggests that the opening angle of the quasar wind is significantly larger than that of the ionization cones, making the coupling of the wind to the gas in the host galaxy more effective and indicating the widespread impact of quasar wind onto its host galaxy. 

Finally, the quasar is powerful enough to contribute to photoionization over most of the host galaxy, but there is one region in the northeast that may be closer to the observer than the redshifted counter-cone, which displays line ratios consistent with star formation. The minimal star formation rate derived from the extinction-corrected \ha\ emission of this region alone is 200 \myr, but it is likely significantly under-estimated since this estimate does not include any obscured star formation nor any star-forming clumps which may be within the ionization cones of the quasar. The northeast star-forming clumps show on average oxygen abundance values (\oh) about half to third solar with a positive metallicity gradient from the center of the brightest star-forming clump towards the outskirts of the star-forming region. Overall the electron density in the northeast clumps is much higher than what is observed in low redshift star-forming galaxies at similar stellar masses, reaching up to 3,000 \eden, with V band extinction up to 3 magnitudes. The overall star-forming conditions are quite different from the local universe but consistent with massive star-forming galaxies at cosmic noon. It is quite possible that the nine clumps we detect across the face of the galaxy are all star-forming, but because of the extra quasar illumination of these clouds, they show up as regions photo-ionized by the quasar on our line-ratio diagrams, and they would be therefore excluded from our star formation rate calculation. Figure \ref{fig:schematic} summarizes all of our results on J1652 presented in this paper.

We are catching the SDSSJ1652 system at a unique time, with ongoing merger activity and quasar-driven outflows that drive shocks and turbulence in the ISM with ongoing vigorous star formation in clumpy star-forming regions across the host galaxy. The unprecedented surface brightness sensitivity of JWST NIRSpec observations in the near-IR allowed for a unique set of discoveries that allowed us to address key questions about the photoionization mechanisms in the quasar host galaxy and in the immediate environment. Evidence for photoionization of the ISM of neighboring galaxies and the surrounding tidal tail on tens of kpc scales strongly supports the scenario that quasar photoionization is a significant contributor to powering the powerful extended Ly$\alpha$ nebula around this source (Gillette et al. in-prep.) and other quasars with similar bolometric luminosities \citep{heck91,Cai19,Arrigoni-Battaia19,OSullivan20}. This indicates that powerful quasars can affect the properties of the ISM of nearby galaxies falling in the ionization cone of the quasar. Direct evidence for the impact of quasar-driven outflows on the ISM in quasar-host galaxies has been a several-decade endeavor. Particularly searching for evidence of shock heating of the ISM that prolongs the lifetime for gas to cool and form stars. The difficulty in detecting this direct impact is largely due to the low surface brightness emission of shock-heated gas at close angular separations to an extremely bright object. Owing to the sensitivity of JWST and the contrast enabled by the NIRSpec IFU and \qfit\ we were able to see the impact of the quasar outflow by detecting clear evidence for shock heating on kpc scales directly in the path of the quasar outflow for the first time at the peak epoch of galaxy and SMBH growth.

\begin{acknowledgments}
A.V., N.L.Z., Y.I., and N.D. are supported in part by NASA through STScI grant JWST-ERS-01335. N.L.Z further acknowledges support by the Institute for Advanced Study through J. Robert Oppenheimer Visiting Professorship and the Bershadsky Fund. We want to thank the anonymous referee for their constructive comments that helped improve the manuscript.

\end{acknowledgments}

\vspace{5mm}
\facilities{JWST(NIRSpec), HST(WFC3) }\\
The data is available at MAST: \dataset[10.17909/qacq-9285]{\doi{10.17909/qacq-9285}}
\software{astropy \citep{Astropy2013, Astropy2018},  reproject \citep{thomas_robitaille_2023_7584411}, \qfit \citep{ifsfit2014}}



\bibliography{bib}{}
\bibliographystyle{aasjournal}

\end{document}